\documentclass[amsmath,amssymb]{revtex4-1}
\usepackage[utf8]{inputenc}
\usepackage{xcolor}
\usepackage{graphicx}
\usepackage{amsmath}
\usepackage{amsfonts}
\usepackage{amssymb}

\begin{document}

\title{Controlling systemic risk - network structures that minimize it and node properties to calculate it}
\author{Sebastian M. Krause}
\affiliation{Division of Theoretical Physics, Rudjer Bo\v{s}kovi\'{c} Institute, Zagreb, Croatia}
\affiliation{Faculty of Physics, University of Duisburg-Essen, Dusiburg, Germany}
\author{Hrvoje \v{S}tefan\v{c}i\'{c}}
\affiliation{Catholic University of Croatia, Ilica 242, 10000 Zagreb, Croatia}
\author{Vinko Zlati\'{c}}
\affiliation{Division of Theoretical Physics, Rudjer Bo\v{s}kovi\'{c} Institute, Zagreb, Croatia}
\author{Guido Caldarelli}
\affiliation{IMT Alti Studi Lucca, Italy}
\affiliation{CNR-ISC, UdR "Sapienza", Rome Italy}
\affiliation{ECLT, Venezia, Italy}

\begin{abstract}
Evaluation of systemic risk in networks of financial institutions in general requires information of inter-institution financial exposures. In the framework of Debt Rank algorithm, we introduce an approximate method of systemic risk evaluation which requires only node properties, such as total assets and liabilities, as inputs. We demonstrate that this approximation captures a large portion of systemic risk measured by Debt Rank. Furthermore, using Monte Carlo simulations, we investigate network structures that can amplify systemic risk. 
Indeed, while no topology in general sense  is {\em a priori} more stable if the market is liquid \cite{roukny2013default}, a larger complexity is detrimental for the overall stability \cite{bardoscia2017pathways}.
Here we find that the measure of scalar assortativity correlates well with level of systemic risk. In particular, network structures with high systemic risk are scalar assortative, meaning that risky banks are mostly exposed to other risky banks. Network structures with low systemic risk are scalar disassortative, with interactions of risky banks with stable banks.     
\end{abstract}

\maketitle

\section{Introduction}

In the past, the stability of the banking sector was mostly analyzed considering measures of the individual banks. Only recently, especially after the 2008 crisis, this has changed. Negative consequences of an interconnected economy and especially of a interconnected financial sector were obvious to the whole world. Furthermore, scientists and policy makers, that were concerned with systemic risk and financial system stability even before the crisis, recognized that there was a serious lack of knowledge on the mechanisms how interconnectedness affects financial stability. For that reason some new analyses including network effects and distress propagation were proposed \cite{battiston2012debtrank,minoiu2013network,chinazzi2013post}. 

Related research direction aimed at expanding the definition of financial system to multi-layer networks, including different assets and valuations, different types of loans etc.~\cite{poledna2015multi,montagna2016multi}.
Applications of these models include central bank regulation \cite{battiston2012debtrank}, individual assessment of systemic risk \cite{poledna2018identifying}, simulations of different policies like for example bank taxation \cite{zlatic2015reduction,poledna2016elimination}.

However, to compute these network risk measures, we need both computer simulations to sample the possible future evolution of the system as well as  a detailed knowledge of the interconnection network of institutions (for example the investments between all pairs of banks \cite{bardoscia2015}). This information about the network is (and just in a few cases) only known to the regulating authorities, for that reason several methods of reconstructing the graph from partial information have been proposed~\cite{cimini2015estimating,cimini2015systemic,cimini2015reconstructing,squartini2018reconstruction,cimini2019statistical}. 
In order to determine the best possible reconstruction, a substantial research analysis has been carried out \cite{anand2017missing}.

Here we present a complementary approach to that of a reconstruction, by showing that a series of risk measures (including network effects) can be understood to a great extent by analyzing the properties of single banks. Indeed the presence of network is taken into account by the choices that managers realize for their institution, as a result close inspection of local (single bank and couple of banks) measures can reveal something about the whole system. Firstly, single bank measures as the interbank leverage (ratio of total investments of a bank into other banks over this banks equity) is enough to understand the first steps of stress propagation (that account for a large part of total stress propagation). Secondly, the  investments between pairs of highly leveraged banks are also increasing stress propagation. We find that investment networks with high systemic risk are highly assortative with respect to single bank risk, while networks with lowest systemic risk are disassortative \cite{newman2003mixing}. 

We use data that are taken from the Italian electronic broker market e-MID (Market for Interbank Deposits) run  by e-MID S.p.A.  ``Societ\`a  Interbancaria per l'Automazione'' (SIA), Milan. The Italian electronic broker Market for Interbank Deposit  (e-MID) covers the entire overnight deposit market in Italy. The information about the parties involved in a transaction allows us to perform risk propagation on real networks as well as a benchmark against which we create artificial networks.

As mentioned before, there is a number of papers which study the risk propagation with Detbrank \cite{battiston2012debtrank}, both for direct application to stress tests \cite{battiston2016leveraging} and to realize a plausible scenario to understand systemic risk \cite{barucca2016network}. Here we follow the approach presented in \cite{bardoscia2015}. This approach simplifies DebtRank method in such a way that one can employ linear algebra, while still preserving the conclusions obtained in other variants of  DebtRank. 

The paper is organized as follows. First, we reintroduce DebtRank algorithm as proposed in \cite{bardoscia2015}. 
Second, we analyze amplification mechanism of the method and rewrite the algorithm in such a way that single node, neighborhood (local), and global contributions to the DebtRank are clearly separated. Third, we propose a Monte Carlo network creation algorithm to test which network configurations are extremal (maximal or minimal) with respect to the DebtRank. Fourth, we present a simple illustrative example, which is followed by empirical results computed from the real data and analytically solvable examples. We finish with analysis of finite size and varying distributions effects on our results presented in previous sections.

\section{Background: Propagating shocks with DebtRank}

Assume $N$ banks, each with equity $E_i$. For every bank $i$, we additionally know, how much it invested in total into other banks. We call this the interbank assets $A_i$ of bank $i$. Additionally we know the liabilities of each bank $i$ to all other banks, called $L_i$. Initially (time $t=0$) we assume no distress, and $\sum_i A_i(0) = \sum_i L_i(0)$. For t=1, we assume external distress on the banks $h(1)$ . According to this distress, the assets $A_i$ have reduced value, as the distressed banks are more likely to bankrupt and therefore not to pay back their debt. On the other hand, liabilities do not get reduced. Here we want to understand network effects of the positive feedback between reduced equity and asset value. For this we follow the DebtRank scenario. More precisely, we are interested in small every-day shocks, where no bank looses all its equity. 

To compute the equity losses, let us assume for the moment we know not only the total amount $A_i$ of assets of bank $i$, but also in which banks $j$ they  invested, denoted with the asset matrix $A_{ij}(0)$. We have $A_i(0)=\sum_j A_{ij}(0)$ and $L_j(0)=\sum_i A_{ij}(0)$. Further we define the matrix $\Lambda$ with elements $\Lambda_{ij}=A_{ij}(0)/E_i(0)$, and the distress parameter $h_i$ describing the relative loss of equity of bank $i$, $h_i(t)=1-E_i(t)/E_i(0)$. According to \cite{bardoscia2015} we have 
\begin{align}
h_i(t) &= h_i(1) + \sum_j \Lambda_{ij} h_j(1) + \sum_j (\Lambda^2)_{ij} h_j(1) + \dots + \sum_j (\Lambda^{t-1})_{ij} h_j(1).\label{eq:h_i_gen}
\end{align}

For a homogeneously distributed initial distress such that $E_i(1)= (1-\psi) E_i(0)$ with a small positive $\psi$ from (\ref{eq:h_i_gen}) we have
\begin{align}
h_i(t)/\psi &= 1 + \sum_j \Lambda_{ij} + \sum_j (\Lambda^2)_{ij} + \dots + \sum_j (\Lambda^{t-1})_{ij}.\label{eq:h_i_psi}
\end{align}
In the remainder of the text we are primarily interested in the total relative systemic equity loss \begin{align}
H(t) &= \sum_i h_i(t) E_i(0) /\sum_j E_j(0)
\end{align}
and especially in its asymptotic value $\lim_{t \rightarrow \infty} H(t) \equiv H^{\infty}$.

\section{Amplification of a small shock hitting all banks}

For a general vector of initial distress $h_i(1)$, the total relative systemic equity loss can be expressed as 
\begin{align}
H^{\infty}=\frac{\sum_i E_i(0) h_i(1)}{\sum_k E_k(0)} + \frac{\sum_i L_i(0) h_i(1)}{\sum_k E_k(0)}+\frac{1}{\sum_k E_k(0)}\sum_{jl} \frac{L_j(0) A_{jl}(0) h_l(1)}{E_j(0)} + {\cal O}(A^3) \, .
\end{align}

As described above, we are interested in a small shock $\psi$ hitting all banks equally, which roughly corresponds to shocks at the macroeconomic level. Although this is necessarily an approximation, it allows us to obtain even more detailed analytical insight into the total relative systemic equity loss using the data on individual banks (node specific data). 
The macroeconomic multiplier $\Psi=H^{\infty}/\psi$ describes, how the external shock is amplified in the banking system. We can rewrite 
\begin{align}
\Psi &= 1 + \frac{\sum_i A_i(0)}{\sum_k E_k(0)} + \frac{\sum_i A_i(0) L_i(0)/E_i(0)}{\sum_k E_k(0)} + \frac{\sum_{ij} A_{ij}(0) L_i(0) A_j(0) /(E_i(0) E_j(0))}{\sum_k E_k(0)}+\Psi^{(\rm res)}\\
 &\equiv 1 + \Psi^{(1)} + \Psi^{(2)} + \Psi^{(3)} + \Psi^{(\rm res)}.
\end{align}
Notice that the terms up to $\Psi^{(2)}$ only depend on the asset and liability sums $A_i$ and $L_i$. 
The term $\Psi^{(3)}$ is the lowest order term including the investment matrix $A_{ij}(0)$. Defining a risk matrix 
\begin{align}
R^{(3)}_{ij} &= L_i(0) A_j(0) /(E_i(0) E_j(0)) \times (1-\delta_{ij}),
\end{align}
$\Psi^{(3)}$ can be written in a more compact way. 
With dimensionless quantities this reads 
\begin{align}
\alpha_{ij} &= A_{ij}(0)/\sum_k E_k(0),\quad a_i=A_i(0)/\sum_k E_k(0),\quad l_i=L_i(0)/\sum_k E_k(0),\quad e_i=E_i(0)/\sum_k E_k(0)\\
\Psi^{(3)} &= \sum_{ij} \alpha_{ij} R^{(3)}_{ij},\quad 
\Psi^{(\rm res)} = \sum_{t=4}^{\infty} \frac{\sum_{ij} E_i(0) (\Lambda^t)_{ij}}{\sum_k E_k(0)}=\sum_{t=4}^{\infty} \sum_{ij} e_i (\Lambda^t)_{ij},\quad \Lambda_{ij}=A_{ij}(0)/E_i(0) = \alpha_{ij} /e_i.
\end{align}
If the eigenvalue of the matrix $\Lambda_{ij}$ with the largest absolute value, in the following called $\lambda$, has the absolute value considerably smaller then one, we can expect the residual term to be a minor correction in $\Psi$. If, on the other hand, $\lambda\geq 1$, the equity loss accelerates infinitely and at least one bank bankrupts. 


Finally, for $t \to \infty$ the relation (\ref{eq:h_i_gen}) can be written at the matrix level as
\begin{align}
h(1)=(I-\Lambda) h^{\infty} \, .\label{eq:inverse}
\end{align}
From the condition that all elements of $h^{\infty}$ are below 1, corresponding to no bankruptcies in the system, it is possible to obtain conditions on initial distress. This result directly reflects the fact that, owing to the network structure encoded in $\Lambda$, the stability of the entire financial network has different sensitivity on the same level of initial distress at various nodes. This information may be of practical importance to financial regulators. In particular, if some $h_i(1)$ is outside allowed range obtained by (\ref{eq:inverse}), regulators should consider intervention, possibly in the form of restructuring the financial network. The practical calculation of $h(1)$ using analytical methods might be prohibitively complicated even for networks of moderate size. A more convenient approach is based on simulations. One can randomly select each component of $h^{\infty}$ in the interval of values corresponding to no bankruptcy ($0 \le h^{\infty}_i < 1$) and calculate $h(1)$ using (\ref{eq:inverse}). With a sufficiently large number of such calculations one can obtain estimates of no-bankruptcy intervals for all components of $h(1)$. This analysis is left for future work.

\section{Minimal and maximal shock amplification $\Psi$}

For understanding the bounds of systemic risk in measures of the shock multiplier $\Psi$, let us minimize or maximize $\Psi(\alpha_{ij})$ by varying $\alpha_{ij}$, given single bank properties $A_i(0)$, $L_i(0)$ and $E_i(0)$. 
We use a stochastic optimization process. For variables $\alpha_{ij}$ we have constraints 
\begin{align}
\sum_j \alpha_{ij} &= a_i,\quad \sum_i \alpha_{ij} = l_j,\quad
\alpha_{ij}\geq 0, \quad \alpha_{ii} = 0.
\end{align}
We want to maximize a more general nonlinear function $F(\alpha_{ij})$ ($F$ will be replaced with $\pm\Psi$ and possible additional terms). 
If we once have a valid matrix $\alpha_{ij}$ fulfilling all constraints, we can add to it a matrix 
\begin{align}
D(i_1,j_1,i_2,j_2)_{ij} &= d \delta_{i,i_1}\delta_{j,j_1} + d \delta_{i,i_2}\delta_{j,j_2} - d \delta_{i,i_1}\delta_{j,j_2} - d \delta_{i,i_2}\delta_{j,j_1},\\
D(0,1,2,3) &=
\begin{bmatrix}
    0       & d & 0 & -d & 0 \\
    0       & 0 & 0 & 0 & 0 \\
    0       & -d & 0 & d & 0 \\
    0       & 0 & 0 & 0 & 0 \\
    0       & 0 & 0 & 0 & 0 \\
\end{bmatrix}.
\end{align}
The example is for $N=5$. To find an initial valid matrix $\alpha_{ij}$, we can start with $\tilde{\alpha}_{ij}=a_i l_j / \sum_k a_k$. We successively subtract matrices of the form $D(i_1,i_1,i_2,i_2)$, until only one diagonal element is left, and further $D(i_1,i_1,i_2,j_2\neq i_2)$ to eliminate the last diagonal element as well. 
For finding matrices $\alpha_{ij}$ with extremal $F(\alpha_{ij})$, we propose updates $\alpha \to \alpha + D(i_1,j_1,i_2,j_2)$, with $D$ involving only off-diagonal elements. If for the updated matrix it would hold $\alpha_{ij}\geq0$, we accept updates with probability
\begin{align}
\min\{1,\exp\{\beta[F(\alpha+D)-F(\alpha)]\}\}.
\end{align}
For $F=\Psi$, $\Psi$ is maximized, while for $F=-\Psi$ it is minimized. The positive parameter $\beta$ regulates, how likely updates away from the optimization goal are accepted. For large $\beta$, such updates are accepted very unlikely. Small $\beta$ can be used to escape local extrema (often combined with an increasing parameter $\beta$ over time, to approach the global extreme in the end of the optimization procedure). In order to force additional constraints for the investment matrix $\alpha_{ij}$, we add further terms 
\begin{align}
F &= \pm \Psi -\beta_{k}\bar{k}(\alpha_{ij})-\beta_{\rm asym}\frac{\sum_{ij}\alpha_{ij} \alpha_{ji}}{\sum_{ij}\alpha_{ij}^2}.
\end{align}
For $\beta_{k}>0$, $\alpha_{ij}$ is more sparse after optimization. The average degree is calculated as $\bar{k}=\bar{k}_{\rm in}=\bar{k}_{\rm out}=\sum_{ij}\Theta(\alpha_{ij})/N$, with theta function $\Theta(x>0)=1$ and $\Theta(0)=0$. With $\beta_{\rm asym}>0$, the investment-matrix is forced to be asymmetric. This has the following meaning: If for a pair of banks $i,j$ it holds $\alpha_{ij}\alpha_{ji}>0$, bank $i$ invests into bank $j$, while at the same time bank $j$ invests into bank $i$. In the e-MID data there is a number of closed loops of length 2, but for the purposes of this paper we chose to suppress them. The reason for this choice is that in overnight market one can easily clear the debt between two parties, and we choose the DebtRank version presented in \cite{bardoscia2015} which does not provide stop in iterations of the DebtRank algorithm. Short loops therefore iterate shock propagation between 2 banks ad infinitum and the correct way to alleviate this problem is to "clear"them into a one directional edge whose weight is the difference between the values of two reciprocal edges.  

\subsection{Illustrative example}

\begin{figure}[htb]
\begin{center}
    \begin{minipage}{0.49\columnwidth}\centering
    (a)\\
    \includegraphics[trim=0 0 0 0 ,clip,width=1.0\columnwidth]{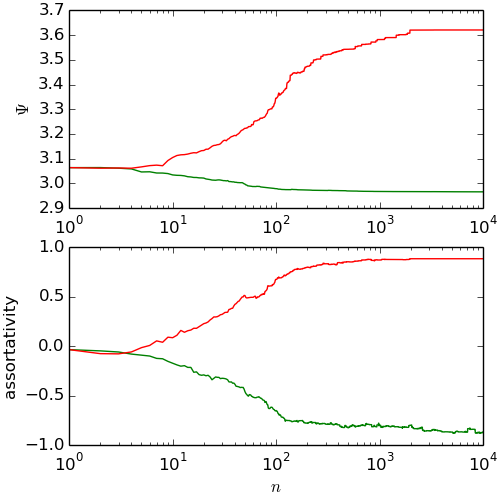}
    \end{minipage}
    \begin{minipage}{0.49\columnwidth}\centering
    (b)\\
    \includegraphics[trim=0 110 0 120 ,clip,width=0.9\columnwidth]{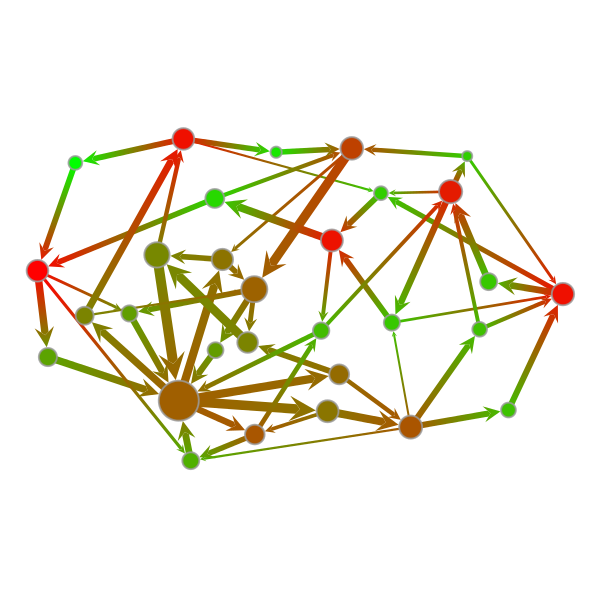}\\
    (c)\\
    \includegraphics[trim=0 160 0 200 ,clip,width=0.9\columnwidth]{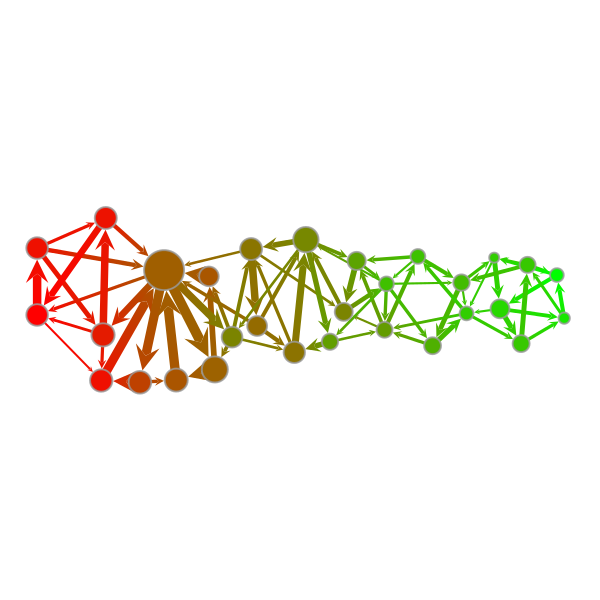}
    \end{minipage}
    \caption{(a) In the upper panel, the shock multiplier $\Psi$ is shown during maximization (red line) and minimization (green line), where $n$ denotes sweeps (with $N^2$ update trials). This is for an artificial example with $N=30$ banks, equities from a Pareto distribution with exponent three, and interbank leverages $0.32<A_i/E_i=L_i/E_i<0.96$ from a uniform distribution. In the lower panel, a scalar assortativity measure with respect to interbank leverage is shown for the same optimization runs. Minimal systemic risk is connected to dissassortative networks (see also (b)), while maximal systemic risk is connected to assortative networks (see also (c)). The networks in (b) and (c) encode the total assets $A_i$ of a bank $i$ as node size, and the interbank leverage $A_i/E_i$ as node color from green (low values) to red (high values).}
    \label{fig:opt}
\end{center}
\end{figure}

For illustration, let us first discuss an artificial example of a network of interbank liabilities. We use a small network with $N=30$ banks, equities from a Pareto distribution with exponent three, and interbank leverages $0.32<A_i/E_i<0.96$ from a uniform distribution. As it is easiest to illustrate and understand the case with $A_i=L_i$, we start with this case. For optimization, we use parameters $\beta=10^6$, $\beta_k=0.1$ and $\beta_{\rm asym}=2.0$. We sum up the first 50 terms of $\Psi$ for assessing update trials, and once a sweep we calculate $\Psi$ using the first 200 terms, with results plotted on the upper panel of fig. \ref{fig:opt}a. The final optimized networks have average degree $\bar{k}=2.0$ (minimization) and $\bar{k}=2.6$ (maximization). Largest eigenvalues are $\lambda=0.67$ (minimization) and $\lambda=0.83$ (maximization). Both connection matrices are strictly asymmetric at the end of optimization. On the lower panel of (a), we see a scalar assortativity measure with respect to interbank leverage $A_i/E_i$.  Correlations among nodes with a scalar property can be described with the assortativity measure $r=[\sum_{xy} x y (e_{xy} - a_x b_y)]/\sigma_a \sigma_b$ \cite{newman2003mixing}. Here $e_{xy}$ is the fraction of links from a vertex of type $x$ to a vertex of type $y$, and node types are assigned with choosing intervals for $A_i/E_i$. Further, we have $a_x=\sum_y e_{xy}$ and $b_y=\sum_x e_{xy}$. We used an implementation provided with graph-tool \cite{peixoto2014graphtool}, where the variance is obtained with the jackknife method. For small systemic risk, highly leveraged banks should both lend from and borrow to banks with small leverage. This is the case for the network with minimized $\Psi$ shown in (b). If highly leveraged banks lend among each other, systemic risk is high, as can be seen in (c).

\subsection{Empirical results}

\begin{figure}[htb]
\begin{center}
    \begin{minipage}{0.35\columnwidth}\centering
    (a)\\
    \includegraphics[trim=0 0 0 0 ,clip,width=1.0\columnwidth]{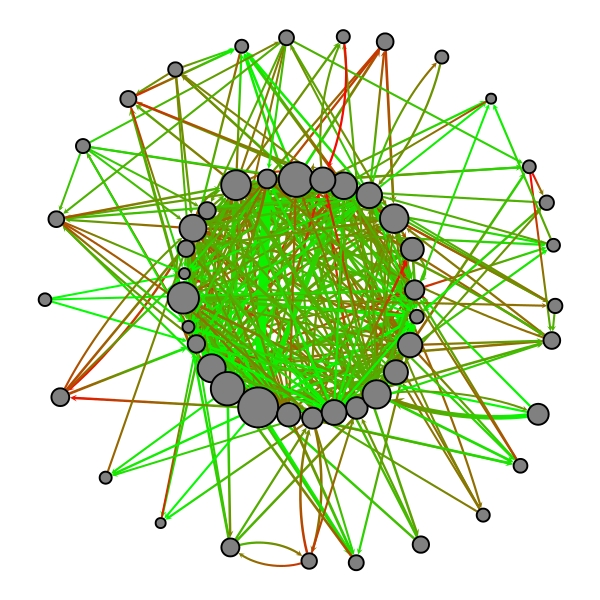}
    \end{minipage}
    \begin{minipage}{0.26\columnwidth}\centering
    (b)\\
    \includegraphics[trim=100 0 100 0 ,clip,width=0.9\columnwidth]{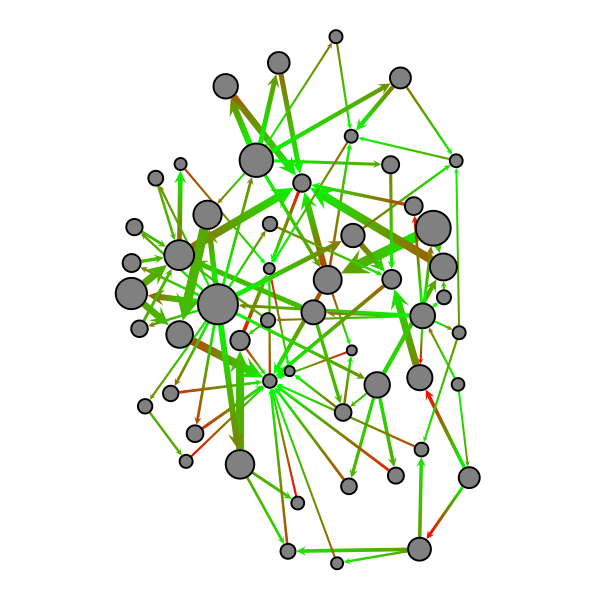}\\
    \end{minipage}
    \begin{minipage}{0.26\columnwidth}\centering
    (c)\\
    \includegraphics[trim=100 0 100 0 ,clip,width=0.9\columnwidth]{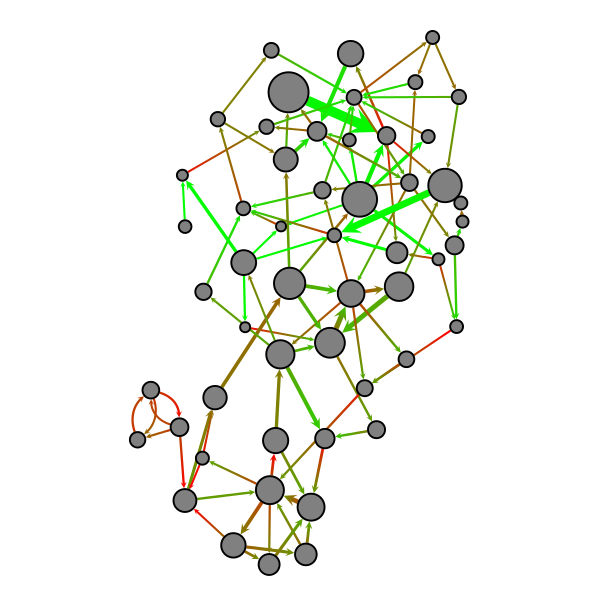}\\
    \end{minipage}\\
    \begin{minipage}{0.38\columnwidth}\centering
    (d)\\
    \includegraphics[trim=0 0 0 0 ,clip,width=1.0\columnwidth]{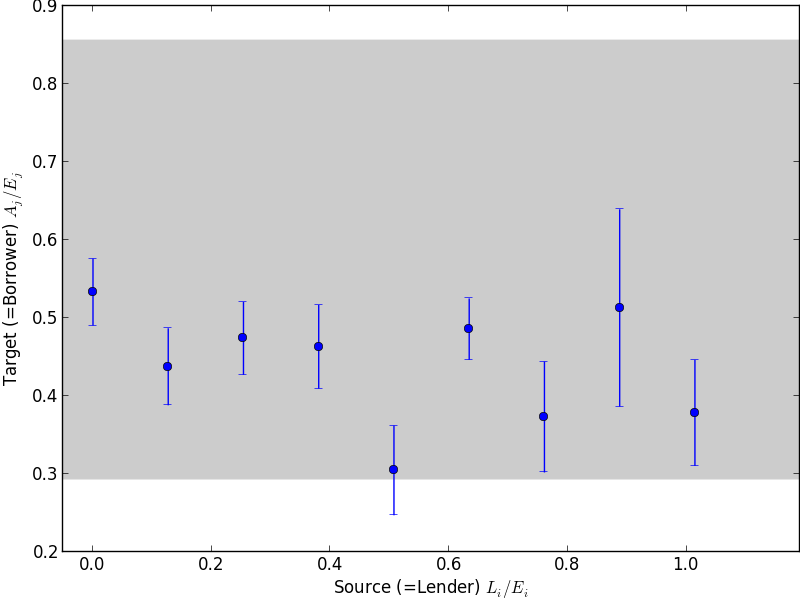}
    \end{minipage}
    \begin{minipage}{0.26\columnwidth}\centering
    (e)\\
    \includegraphics[trim=0 0 0 0 ,clip,width=1.0\columnwidth]{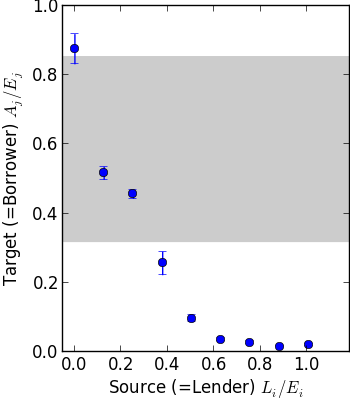}
    \end{minipage}
    \begin{minipage}{0.26\columnwidth}\centering
    (f)\\
    \includegraphics[trim=0 0 0 0 ,clip,width=1.0\columnwidth]{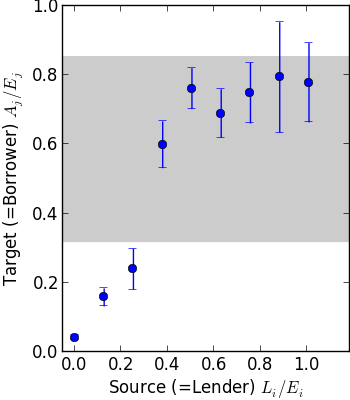}
    \end{minipage}
    \caption{(a) Liability network with $N=53$ banks in the Italian market in 1999. (d) For this network, source and target properties are uncorrelated. (e) After minimizing systemic risk,  the network becomes disassortative, with anti-correlations among different scalar properties for source and target node. For $A_i=L_i$, these correlations can be described with the simpler measure of scalar disassortativity as shown in fig. \ref{fig:opt}a on the bottom. The gray band indicates area between first and third quartile for $A_i/E_i$, so half of values $A_i/E_i$ around the median lies within the gray area. We see that for small systemic risk, a bank $i$ with high $L_i/E_i$ should lend to a bank $j$ with low $A_j/E_j$. Interpretation: A bank $i$ with high $L_i/E_i$ has high impact on its lenders, while a bank $j$ with low $A_j/E_j$ has only small exposure to its borrowers, thus shocks are dampened. (b) 
    The network with minimal risk. Total assets $A_i$ of a bank $i$ shown as node size, and the interbank $L_i/E_i$ as edge color at the edge source, $A_j/E_j$ at edge target, color from green (low values) to red (high values). (c) Network with maximal risk. (f) Maximal systemic risk  is connected to assortativity. }
    \label{fig:opt2}
\end{center}
\end{figure}

We use an interbank liability data-set for the European market involving Italian banks in the year 1999. For a shock in the night before the last trading day in July, Friday July 30. 1999, we consider all outstanding liabilities with lifetime at least the next five trading days. These contracts thus have to be repaid earliest the upcoming Friday after one week. This choice is to guarantee that shock propagation due to devaluation of contracts has time to take place, what is in question for overnight obligations. Possible contract durations are thus starting from two weeks, up to one year. We construct the network of all 218 involved banks, and reduce it to the largest strongly connected component, including $N=53$ banks. As the data-set is anonymized, we have to reconstruct the equity of the banks. We choose $E_i={\rm max}(A_i,L_i)\times 1.25 \times \xi_i$ with $\xi_i$ from a normal distribution with mean one and standard deviation $0.2$. The resulting network can be seen in fig. \ref{fig:opt2}a. Total assets $A_i$ of a bank $i$ shown as node size, and the interbank $L_i/E_i$ as edge color at the edge source, $A_j/E_j$ at edge target, color from green (low values) to red (high values). We found a shock amplifier $\Psi=1.90$ for this network. In (d) we analyze for this network correlations between lenders liabilities divided by equity (source $L_i/E_i$) and borrowers leverage (target $A_i/E_i$). The average $A_i/E_i$ of all target nodes is plotted which are reached from source nodes with values $L_i/E_i$ from a certain interval. We see that there are no significant correlations between lenders liabilities divided by equities and borrowers leverage. For $A_i=L_i$, these correlations simplify and can be described with the scalar assortativity as shown in fig. \ref{fig:opt}a on the bottom. 

The network consists of 763 edges among the 53 banks, therefore the average degree is 14.4. Assets $A_i$ and liabilities $L_i$ are mildly correlated with a pearson correlation of 0.11. In total 36 of the directed edges have a counter part in the opposite direction, so some loops of length two are present. The largest $A_i$ is 875 million Euros, the largest $L_i$ is 1132 million Euros. All assets sum up to 7.04 billion Euros, so do the liabilities. Using 100 different samples of equities $E_i$ we found $\langle\Psi\rangle=1.87$ with standard deviation $0.06$. 

As for the illustrative example, we minimize and maximize the shock amplifier $\Psi$ with final sweep $n=10^4$, $\beta=10^6$, $\beta_k=0.1$ and $\beta_{\rm asym}=2.0$. We sum up the first 50 terms of $\Psi$ for assessing update trials. After the final sweep we calculate $\Psi$ using the first 200 terms, with results $\Psi=1.80$ (minimization) and $\Psi=2.25$ (maximization). The final optimized networks both have average degree $\bar{k}=2.0$. The connection matrix minimizing shock amplification is strictly asymmetric at the end of optimization, while for maximization we see a small number of loops of length two. Results are shown in fig. \ref{fig:opt2}. With (b) and (e) we see that an investment matrix with minimized $\Psi$ has a more subtle kind of scalar disassortativity, as compared to the illustrative example with $A_i/E_i=L_i/E_i$. Here, systemic risk is minimized, when banks with high $L_i/E_i$ lend to banks with low $A_i/E_i$. With (c) and (f), we see that the network of the maximized systemic risk is assortative. It is important to stress that this structure is very different from the typical core-periphery structure usually observed in financial networks \cite{fricke2015core,rombach2017core}. It is also important to stress that risk minimization in principle reduces the number of edges in the network, therefore reducing the risk diversification of single financial institution. The apparent paradox was previously addressed in \cite{battiston2012liaisons}. We also have to stress that scalar assortativity is very different from network assortativity. Previous analysis of cascades in complex networks \cite{d2012robustness}, showed that cascades are (analogous to systemic risk spreading) inhibited by \emph{network assortative} structures, while this analysis shows that systemic risk is amplified with \emph{scalar assortativity}. This results are not opposed to each other but are complementary to each other.   

\subsection{Analytically solvable examples}

There is another strong indicator, why correlations between source $L_i/E_i$ and target $A_j/E_j$ are dominating in the optimization: For constant $C=L_i/E_i$ or constant $C=A_i/E_i$ (e.g. no positive or negative correlations possible), $\Psi$ is constant, independent of the investment matrix $A_{ij}$. Let us first show this for $C=A_i/E_i$. The terms up to $\Psi^{(2)}$ are anyhow independent of $A_{ij}$. For higher terms we can write $\Psi^{(3)}+\Psi^{({\rm res})}=\sum_{t=3}^{\infty} \sum_{ij} e_i (\Lambda^t)_{ij}$. We can define a stochastic matrix with elements $S_{ij}=A_{ij}/E_i C$, as $A_i/E_i=\sum_j A_{ij}/E_i=C$. With $\sum_j S_{ij}=1$ and $\Lambda_{ij}=S_{ij}C$, we have 
\begin{align}
\Psi^{(3)}+\Psi^{({\rm res})} &= \sum_{t=3}^{\infty} \sum_{ij}e_i C^t (S^t)_{ij}= \sum_{t=3}^{\infty} C^t \sum_i e_i = \sum_{t=3}^{\infty} C^t.
\end{align}
Here we use properties of stochastic matrices, $\sum_{j} (S^2)_{ij}=\sum_{jk} S_{ik} S_{kj}=1$ etc. For liability sums being constant $C=L_i/E_i$, we can define a stochastic matrix $S_{ij}=A_{ij}/E_j C$, here with $\sum_i S_{ij}=1$. We have $\Lambda_{ij}=S_{ij} C E_j/E_i$, and $\sum_{ijkl} E_i \Lambda_{ij} \Lambda_{jk} \Lambda_{kl}=C^3 \sum_{ijkl} S_{ij} S_{jk} S_{kl} E_l=C^3 \sum_l E_l$, with the same result $\Psi^{(3)}=C^3$ as for constant leverage. The same holds for higher terms. With this finding, other more subtle properties of the investment matrix, as second neighbor correlations, can only play a limited role. Further, we found an approximation for banks with interbank leverage from a sharply peaked distribution (${\rm max}_i |A_i/E_i-\left<A_j/E_j\right>_j|\ll \left<A_i/E_i\right>_i$). In this case, the macroscopic shock amplification is mostly independent of the investment network and a simple function of the average leverage $\Psi \approx \sum_{t=0}^{\infty} (\left<A_i/E_i\right>_i)^t=1/(1-\left<A_i/E_i\right>_i)$, with geometric sum only for $\left<A_i/E_i\right>_i<1$. 

Let us now discuss a case, where the optimization of $\Psi$ can be performed explicitly. We have $N=n_1+n_2$ banks with identical equity $E_i=E$, $e_i=1/N$. With this choice, we have $\Lambda_{ij}=A_{ij}/E$. The first $n_1$ banks have $A_i/E=L_i/E=c_1$, while the last $n_2$ banks are less leveraged with $A_i/E=L_i/E=c_2<c_1$. Illustrated for $n_1=2$ and $n_2=3$, let us introduce the following parametrized matrix 
\begin{align}
\Lambda &=
\begin{bmatrix}
    \frac{c_1}{n_1}       & \frac{c_1}{n_1} & 0 & 0 & 0 \\
    \frac{c_1}{n_1}       & \frac{c_1}{n_1} & 0 & 0 & 0 \\
    0       & 0 & \frac{c_2}{n_2} & \frac{c_2}{n_2} & \frac{c_2}{n_2} \\
    0       & 0 & \frac{c_2}{n_2} & \frac{c_2}{n_2} & \frac{c_2}{n_2} \\
    0       & 0 & \frac{c_2}{n_2} & \frac{c_2}{n_2} & \frac{c_2}{n_2} \\
\end{bmatrix}+\kappa
\begin{bmatrix}
    -\frac{n_2}{n_1}       & -\frac{n_2}{n_1} & 1 & 1 & 1 \\
    -\frac{n_2}{n_1}       & -\frac{n_2}{n_1} & 1 & 1 & 1 \\
    1       & 1 & -\frac{n_1}{n_2} & -\frac{n_1}{n_2} & -\frac{n_1}{n_2}  \\
    1       & 1 & -\frac{n_1}{n_2} & -\frac{n_1}{n_2} & -\frac{n_1}{n_2}  \\
    1       & 1 & -\frac{n_1}{n_2} & -\frac{n_1}{n_2} & -\frac{n_1}{n_2}  \\
\end{bmatrix}
= \Lambda_{a}+\kappa \Delta.
\end{align}
The matrix $\Lambda_{\rm a}$ is maximally assortative, as only banks of the same type interact. For a simpler notation, we allow for self-links. The diagonal elements can easily be emptied into links among banks of the same type. This keeps $\Psi$ unchanged. With $\Lambda_{ij}\geq 0$, we have $0\leq \kappa \leq {\rm min}(c_1/n_2,c_2/n_1)$. For the maximal value of $\kappa$, the two bank types interact as much as the constraints allow. Therefore, this is the maximally dissassortative case. The change in $\Psi$ for an infinitessimal increase of dissassortativity, going from $\Lambda$ to $\Lambda+{\rm d}\kappa \Delta$, is 
\begin{align}
{\rm d}\Psi &= \sum_{t=3}^{\infty} \sum_{p=2}^{t-1} \sum_{ij} (\Lambda^{p} \Delta \Lambda^{t-p})_{ij} {\rm d}\kappa/N,\\ 
\sum_{ij} (\Lambda^{p} \Delta \Lambda^{t-p})_{ij} &=- f(\Lambda^p) f(\Lambda^{t-p})\quad {\rm with }\quad f(\Lambda^p) = n_1 (\Lambda^p)_{11}+(n_2-n_1) (\Lambda^p)_{1N} - n_2 (\Lambda^p)_{NN}.
\end{align}
We neglect higher order terms in ${\rm d}\kappa$ and use the fact that $\sum_i \Delta_{ij}=0$, such that this matrix only occurs in between matrices $\Lambda$. With showing that $\sum_{ij} (\Lambda^{p} \Delta \Lambda^{t-p})_{ij}\leq 0$ for all $0<p<t$, we show that the most assortative connection matrix implies largest shock propagation, while the most dissassortative matrix implies smallest shock propagation. We show that $f(\Lambda^p)=f(\Lambda) C_p$ with $C_p$ positive. Using $(\Lambda^p)_{1N}=n_1 \Lambda_{11}(\Lambda^{p-1})_{1N}+n_2 \Lambda_{1N}(\Lambda^{p-1})_{NN}$ and analog expressions for $(\Lambda^p)_{11}$ and $(\Lambda^p)_{NN}$, we can write $f(\Lambda^p)=n_1 (\Lambda^{p-1})_{11} f(\Lambda)+n_2 \Lambda_{NN} f(\Lambda^{p-1})$. This is a positive multiple of $f(\Lambda)$, if this holds for $f(\Lambda^{p-1})$. With the condition being trivially fulfilled for $f(\Lambda^1)$, we can use induction to prove it for any $p$. 

For $\Lambda=\Lambda_{\rm ass}$, the simple closed form result $\Psi=\sum_{t=0}^{\infty} c_1^t + c_2^t$ holds. The dominating term $c_1^t$ grows or shrinks exponentially with $t$. The minimized $\Psi$ is a lengthy polynomial in $c_1$, $c_2$, $n_1$ and $n_2$ which cannot be easily reduced into a closed form expression. With an ansatz $v=(1,1,\dots,a,a,\dots)$ for the eigenvector with largest eigenvalue $\lambda$, we find 
\begin{align}
\lambda &= \frac{c_1-\kappa n_2+c_2 -\kappa n_1}{2}+\left\{\frac{[c_1-\kappa n_2-(c_2 -\kappa n_1)]^2}{4} + \kappa^2 n_1 n_2\right\}^{1/2}.
\end{align}
Assume many healthy banks and a few highly leveraged banks: $n_1=5$, $c_1=2$, $n_2=50$, $c_2=0.5$. For $\Lambda_{\rm a}$ we have $\lambda=c_1=2$ with the first $n_1$ banks going bankrupt. For largest possible $\kappa$, we have $\lambda=0.8$. Here all banks survive a small macroeconomic shock. In this latter case, the first $n_1$ banks do not lend among each other, and the healthy banks dedicate a share of $2/5$ for interactions with the first $n_1$ banks and remain a share of $3/5$ for interactions among each other.

\subsection{Finite size effects and varying distributions of single bank properties}

\begin{figure}[htb]
\begin{center}
    \includegraphics[trim=0 0 0 0 ,clip,width=1.0\columnwidth]{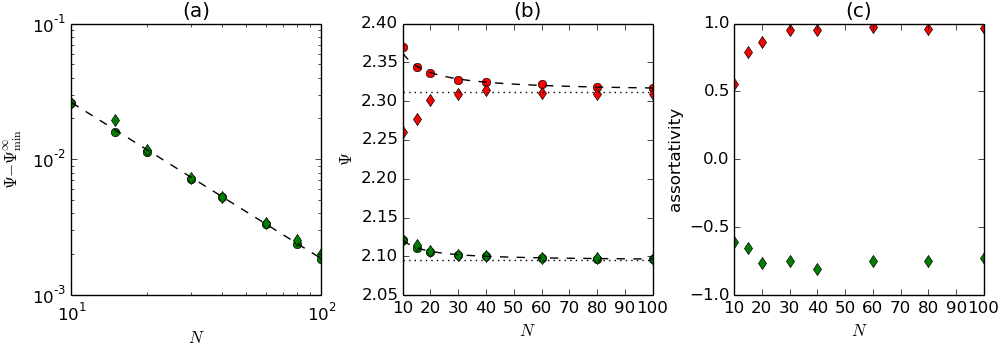}
    \caption{Finite size effects for unrestricted optimization (circles) and restricted optimization, where small degree and asymmetric investment-matrix is forced (diamonds). (a) With finite size scaling we find that for large $N$, $\Psi$ after minimization approaches $\Psi_{\rm min}^{\infty}\approx 2.0947\pm 5\times 10^{-4}$, with a finite size deviation about $\propto N^{-1.15}$. Numerical results are shown with circles (unrestricted) and diamonds (restricted optimization). The dashed line indicates a power law with exponent -1.15. (b) Results of (a) are repeated with linear scale (green symbols and lower dashed line), and compared to results of maximization (red symbols and upper dashed line indicating results of a finite size scaling). The dotted lines indicate $\Psi_{\rm min}^{\infty}$ and $\Psi_{\rm max}^{\infty}\approx 2.315\pm 0.01$. (c) Assortativity after restricted optimization for minimization (green diamonds) and maximization (red diamonds). Results indicate that, independent of the network size, least risky networks are strongly disassortative, while most risky networks are strongly assortative with respect to leverage. }
    \label{fig:scaling}
\end{center}
\end{figure}

For discussing finite size effects with varying network size $N$, we choose $E_i=1$ for all banks, and $A_i/E_i=L_i/E_i=0.2+0.6 i/(N-1)$. This way, single bank properties for networks of different size are similar, and there is no need to average over many realizations of single bank properties. We optimize for $n_{\rm max}=5 \times 10^3$ sweeps with increasing parameter $\beta=10\times N^2 \times 100^{n/n_{\rm max}}$. The algorithmic cost per optimization sweep scales with $N^4$, as for every microscopic update trial, matrix multiplications have to be performed (scaling with $N^2$), and there are $N^2$ microscopic update trials in a sweep. With a choice of small values for leverage, we can use only the first 13 terms in $\Psi$ for assessing update trials. Final $\Psi$ is calculated with 103 terms. Unrestricted optimization is performed with $\beta_k=\beta_{\rm asym}=0$, results with restrictions are found using $\beta_k=0.1$ and $\beta_{\rm asym}=2.0$. In fig. \ref{fig:scaling}a we see a finite size scaling for unrestricted (green circles) and restricted (green diamonds) minimization. We found an asymptotic result $\Psi_{\rm min}^{\infty}\approx 2.0947\pm 5\times 10^{-4}$, and $\Psi-\Psi_{\rm min}^{\infty}\propto N^{-1.15}$. We also performed a finite size scaling for results of unrestricted maximization of $\Psi$ (not shown). This has less convincing results, indicating that local maxima are a problem. We found $\Psi_{\rm max}^{\infty}\approx 2.315\pm 0.01$. In (b) we see results for minimization (green) and maximization (red). For small networks, restricted maximization results (red diamonds) are far below the unrestricted case (red circles). However, deviations are small for larger networks. In (c) we see that networks with maximized $\Psi$ are strongly assortative, while networks with minimized $\Psi$ are strongly disassortative.

\begin{figure}[htb]
\begin{center}
    \begin{minipage}{0.48\columnwidth}\centering
    \includegraphics[trim=0 189 0 0 ,clip,width=1.0\columnwidth]{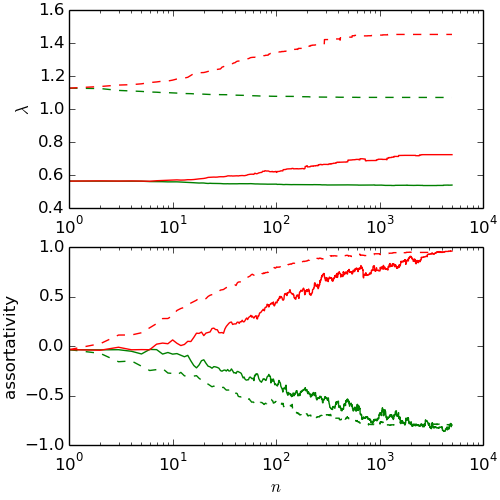}\\
    $n$
    \end{minipage}
    \begin{minipage}{0.48\columnwidth}\centering
    \includegraphics[trim=0 0 0 170 ,clip,width=1.0\columnwidth]{finance_graph_opt_psi_data_30_.png}
    \end{minipage}
    \caption{Rescaling single bank leverage $A_i/E_i \to c\times A_i/E_i$, stress propagation can switch from dampened to exponentially growing. We see this for an example case with $c=2$. While the largest eigenvalue $\lambda$ of the stress propagation matrix $\Lambda$ increases from below one (solid lines, minimization green, maximization red) to above one (dashed lines), the optimization procedure has a similar outcome with respect to assortativity in both cases.}
    \label{fig:distributions}
\end{center}
\end{figure}

We already discussed, how correlations among single bank properties $A_i/E_i$ and $L_i/E_i$ affect results. Let us now discuss the outcome with rescaling $A_i/E_i \to c \times A_i/E_i$ and $L_i/E_i \to c \times L_i/E_i$. We choose $E_i=1$ for all banks, and $A_i/E_i=L_i/E_i=c\times (0.2+0.6 i/(N-1))$, with $N=30$ banks. In fig. \ref{fig:distributions} we see results for $c=1$ (solid lines) and $c=2$ (dashed lines). We optimize for $n_{\rm max}=5 \times 10^3$ sweeps with increasing parameter $\beta=10\times N^2 \times 100^{n/n_{\rm max}}$. As the losses grow exponentially for $c=2$, we only use the first 6 terms in $\Psi$ for assessing update trials. On the left of the figure, we see the largest eigenvalue $\lambda$ of the stress propagation matrix $\Lambda$ during optimization. $\lambda$ is larger then one for $c=2$ (dashed lines). This means that even a very small initial shock causes an exponentially growing stress propagation, finally causing at least one bankrupt bank. On the right of the figure, we see that monitoring assortativity while optimization indicates a similar behavior, even if stress propagation changes from dampened (solid lines) to exponentially growing (dashed lines).

\section{Summary and outlook}

We saw that in the framework of DebtRank, most risky investment networks are highly assortative with respect to lenders liabilities divided by equity (source $L_i/E_i$) and borrowers leverage (target $A_i/E_i$). We tested this for artificial samples of single bank properties, finding that the effect is robust regarding to correlations among single bank properties, network size and finally, it is also a common feature of dampened or exponentially growing stress propagation. Also for empirical data we found this behavior. Finally we performed the optimization analytically for a network with two types of banks.

Two main results of this paper are: (i) shock propagation in financial networks can be approximately calculated from single node properties only and (ii) this shock propagation can be minimized by making financial networks disassortative. Besides an obvious advantage in (i) that the computation using single node properties only is simpler and faster, there exist other advantages and potential applications of these results.

The possibility to (approximately) estimate shock amplification in interbank networks from single bank properties brings additional advantage for financial regulators. Namely, single bank properties necessary for such estimation, such as their total assets, $A_i$, and liabilities $L_i$ are cumulative quantities and, as such, they change more slowly than changes in the structure of the interbank networks. In particular, on a daily basis, we do not expect total assets or liabilities of the bank to change significantly. However, it is reasonable to expect that at the same daily timescale any bank in the network would engage in lending to or borrowing from many new banks, or changing amount of lending/borrowing for other banks that the said bank is already connected to. Thus, in the regime where the approximation of shock propagation is reliable using only first terms that depend on single bank properties, these estimates are also expected to remain reliable for as long as these single bank properties do not change significantly, and much longer than the typical scale on which the interbank network changes.

The association of scalar disassortative network structures with lower systemic risk, gives regulators more ''degrees of freedom" in resolving situations where vulnerability of a small number of banks threatens the entire network. Namely, there are many network structures with high disassortativity and it is easier for regulators to find or realize one of them if realistic legal, liquidity or even political constraints exist.

An interesting parallel with physical systems also arises from this analysis. Namely, if we classify leverage in to discreet categories, than we can possibly map them to spin systems like Potts model. If this analogy holds, one could associate low risk structures with a variant of antiferro Potts model, while networks which exhibit more risk could possibly be associated with ferro variant of Potts model. Weather this analogy holds is beyond the scope of this paper, but if the mapping of systemic risk model to such a well studied statistical physics model would be obtained, a community of scientists that study systemic risk could greatly benefit from accumulated knowledge.

Finally, the approach of estimating shock propagation in interbank networks from single bank properties only, provides a novel possibility for public oversight of financial system stability. As banks publish public financial statements in regular intervals, and these statements contain data on total borrowing from or lending to other banks in the financial system, it is in principle possible for anyone to compute the lower bound on the systemic risk for various scenarios of initial financial distress. In this way, monitoring systemic risk in the financial system would no longer be limited to regulatory authorities. 

\section{Acknowledgments}

H\v{S} and VZ had their research supported by the European Regional Development Fund under the grant KK.01.1.1.01.0009 (DATACROSS). VZ was supported by QuantiXLie Center of Excellence, a project cofinanced by the Croatian Government and European Union through the European Regional Development Fund— the Competitiveness and Cohesion Operational Programme (Grant KK.01.1.1.01.0004). VZ was also supported by the European Union through the European Regional Development Fund - the Competitiveness and Cohesion Operational Programme (KK.01.1.1.06), and by the H2020 CSA Twinning project No. 692194, RBI-T-WINNING.

\bibliography{ControllingSystemicRisk}

\end{document}